\documentclass[floatfix,amsmath,amssymb]{revtex4}
\usepackage{epsfig}

\usepackage[latin1]{inputenc}
\begin{document}
\title{Time evolution of the energy density inside a non-static cavity with
a thermal, coherent and a Schrödinger cat state}

\author{D T Alves$^{1}$, E R Granhen$^{2}$, M G Lima$^{1}$, H O Silva$^{1}$ and 
A L C Rego$^{1}$}

\address{
(1) - Faculdade de F\'\i sica, Universidade Federal do
Par\'a, 66075-110, Bel\'em, PA,  Brazil
\\
(2) - Centro Brasileiro de Pesquisas Físicas, Rua Dr. Xavier Sigaud,
150, 22290-180, Rio de Janeiro, RJ, Brazil}

\begin{abstract}
In this paper we investigate the time evolution of the energy density for a real massless scalar field in a two-dimensional spacetime, inside a non-static cavity, taking as basis the exact numerical approach purposed by Cole and Schieve.
Considering Neumann and Dirichlet boundary conditions, we investigate
the following initial states of the field: vacuum, thermal state, the coherent state and the Schrödinger cat state. 
\end{abstract}

\maketitle
\section{Introduction}
\hspace{0.5 cm}
Moore \cite{Moore-1970}, in the context of a real massless scalar field inside a 
non-static cavity, obtained an exact formula for the expected value of the energy-momentum tensor,
given in terms of a functional equation, usually called Moore's equation.
For this equation there is no general technique of analytical solution, but
exact analytical solutions for particular movements of the boundary \cite{Fulling-Davies-PRS-1976-I,
Law-PRL-94,solucoes-analiticas-exatas}, 
and also approximate analytical solutions (see, for instance \cite{Dodonov-JMP-1993,Dalvit-PRA-1998}), have been obtained. 
On the other hand, Cole and Schieve \cite{Cole-Schieve-1995} 
proposed a numerical method to solve exactly the Moore equation for a general law of motion
of the boundary. Moreover, the dynamical Casimir effect has been investigated for a thermal bath as the initial field state in a non-static cavity \cite{Dodonov-JMP-1993,temperatura-uma-fronteira,plunien-PRL-2000,alves-granhen-lima-PRD-2008,temperatura-cavidade},
and also for a coherent state \cite{alves-granhen-lima-PRD-2008, Alves-Farina-Maia-Neto-JPA-2003, estados-coerentes,estados-coerentes-2}.
Squeezed states have also been considered \cite{alves-granhen-lima-PRD-2008}.

In the present paper, taking in account a technique of calculation based on
the Cole-Schieve approach \cite{Cole-Schieve-1995,Cole-Schieve-2001}, we examine the energy density in a non-static cavity
for the following initial states of the field: vacuum, thermal state, coherent state and the Schrödinger cat state. 
The Cole-Schieve approach is appropriate, since we consider a certain non-oscillatory law of motion, with large displacements and
relativistic velocities. In this sense, perturbative techniques requesting non-relativistic oscillating movements with small
amplitudes are not suitable in this case. The article is organized as follows: in Sec. 2 we show the general field solution.
In Sec. 3 we investigate the non-static cavity for a vacuum and a thermal initial field states.
In Sec. 4 we analyze the coherent and Schrödinger cat states as the initial field states. 
In Sec. 5 we make some final comments.

\section{General field solution}
\hspace{0.5 cm}
Let us start considering the field satisfying the Klein-Gordon 
(we assume throughout this paper $\hbar=c=k_B=1$) and obeying conditions imposed at the static boundary located at $x=0$,
and also at the moving boundary's position at $x=L(t)$, where $x=L(t)$ is a prescribed law
for the moving boundary and $L(t<0)=L_0$, with $L_0$ being the length of the cavity in the static situation.
We consider four types of boundary conditions. 
The Dirichlet-Neumann (DN) boundary condition imposes Dirichlet condition at the static boundary,
whereas the space derivative of the field
taken in the instantaneously co-moving Lorentz frame 
vanishing (Neumann condition) at the moving boundary's position:
We also consider: Dirichlet-Dirichlet (DD), Neumann-Neumann (NN) and Neumann-Dirichlet (ND)
boundary conditions. A general solution of the wave equation can be written as:
\begin{equation}
\hat{\psi}(t,x)=\lambda(\hat{A}+\hat{B}\psi^{(0)})+\sum^{\infty }_{n=1-2\beta}\left[
\hat{{a}}_{n}\psi_{n}\left( t,x\right) +H.c.\right],
\label{field-solution-1}
\end{equation} 
where the field modes $\psi_{n}(t,x)$ are given by
\begin{eqnarray}
\psi_{n}(t,x)=\frac{1}{\sqrt{4(n+\beta)\pi}}\left[\gamma e^{-i(n+\beta)\pi R(v)}+\gamma ^{*} e^{-i(n+\beta)\pi R(u)}\right],
\label{field-solution-2}
\end{eqnarray} 
with $\psi^{(0)}=\left[R(v)+R(u)\right]/2$ \cite{Dalvit-JPA-2006}, $u=t-x$, $v=t+x$, 
and $R$ satisfying the functional equation
$
R[t+L(t)]-R[t-L(t)]=2\;,
$
which is the Moore equation. The operators
$\hat{A}$ and $\hat{B}$ satisfy the commutation rules 
$\left[\hat{A},\hat{B}\right]=i$, 
$\left[\hat{A},\hat{a}_{n}\right]=\left[\hat{B},\hat{a}_{n}\right]=0$.    
The NN solution is recovered for $\lambda=\gamma=1$ and $\beta=0$. The other three cases are recovered if $\lambda=0$ and: $\beta=0$ and $\gamma=i$ for the DD case;
$\beta=1/2$ and $\gamma=i$ for the DN case; $\beta=1/2$ and  $\gamma=1$ for the ND case.
%
\section{Diagonal states: vacuum and thermal states} 
\hspace{0.5 cm}
As examples of initial field states such that the density matrix is diagonal in the Fock basis, let
us consider the vacuum and the thermal state. It can be shown that the expected value of the energy density operator ${\cal T}=\langle\hat{T}_{00}(t,x)\rangle$ can be split in ${\cal T}={\cal T}_{\footnotesize\;\mbox{vac}}+ {\cal T}_{\footnotesize\;\mbox{non-vac}}$, where ${\cal T}_{\footnotesize\;\mbox{vac}}$ is the contribution to the energy density due the vacuum part and ${\cal T}_{\footnotesize\;\mbox{non-vac}}$ is the non-vacuum contribution due to the real particles in the initial state of
the field. Hereafter we consider the averages $\langle...\rangle$ taken over initial field states 
annihilated by $\hat{B}$.
Let us start considering the vacuum as the initial field state (${\cal T}_{\footnotesize\;\mbox{non-vac}}=0$). 
The vacuum contribution to the energy density inside the oscillating cavity can be written as \cite{Moore-1970,Fulling-Davies-PRS-1976-I}
${\cal T}_{\footnotesize\;\mbox{vac}} = -f(v) -f(u), 
\label{T-vac-ren}
$
where:
\begin{equation}
f=\frac{|\gamma|^2}{24\pi}\left\{\frac{R^{\prime\prime\prime}}{R^{\prime}}-\frac{3}
{2}\left(\frac{R^{\prime\prime}}{R^{\prime}}\right)^{2}+\pi
^{2}\left[\frac{1}{2}-3(\beta-\beta^2)\right]{R^{\prime}}^{2}\right\}. 
\label{T-vac-f}
\end{equation}
For the static situation the function $R$ is given by $R(z)={z}/{L_0}$ and its first derivative is a constant ${R}^{'}(z)={1}/{L_0}$.
From this equation we get the known static Casimir energy densities: ${\cal T}_{\footnotesize\;\mbox{vac}}^{\footnotesize\;\mbox{DD}}={\cal T}_{\footnotesize\;\mbox{vac}}^{\footnotesize\;\mbox{NN}}=-\pi/(24L_0^2)$ \cite{Moore-1970,Fulling-Davies-PRS-1976-I} and
${\cal T}_{\footnotesize\;\mbox{vac}}^{\footnotesize\;\mbox{DN}}={\cal T}_{\footnotesize\;\mbox{vac}}^{\footnotesize\;\mbox{ND}}
=\pi/(48L_0^2)$ \cite{Boyer-AJP-2003}, where the superscripts DD, NN, DN and ND means the types of boundary conditions considered in the calculations.

\begin{figure}
\begin{center}
\epsfig{file=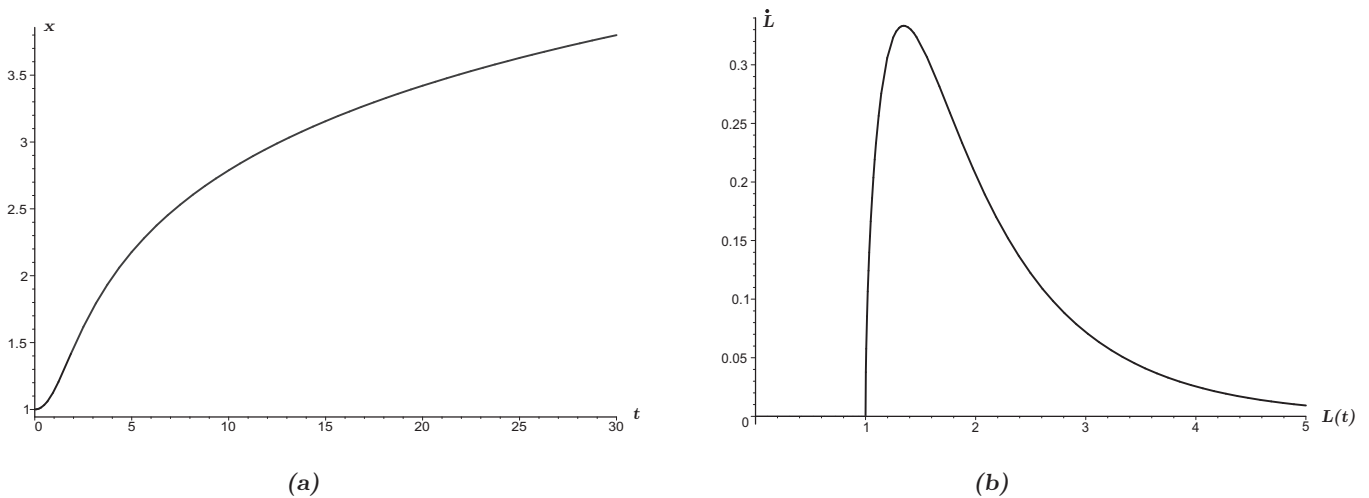,angle=-00,width=1.0\linewidth,clip=}
\caption{\footnotesize {\bf(a)} Moving mirror trajectory defined by Eq. (\ref{trajetoria_walker}); {\bf(b)} Moving
mirror velocity (vertical axis) as function of the boundary position.}
\label{velocidade-trajetoria-WD}
\end{center}
\end{figure}

For oscillatory laws of motion,
the energy density and the force acting on the moving boundary have been considered in the literature.
Here, we investigate the following particular trajectory $x=L(t)$,
analogous to the one proposed by Walker and Davies \cite{Walker-Davies-1982}:
\begin{equation}
t=B-L+A\left( e^{-2(B-L)/B}-1\right) ^{1/2},  
\label{trajetoria_walker}
\end{equation}
where we assume $L(0)=L_0=1$, $A$ and $B$ are constants, with $A>B$ so that $|\dot{L}| <1$. This is a smooth and asymptotically static trajectory at $t\rightarrow \infty $ (see Fig. \ref{velocidade-trajetoria-WD} (a)). The mirror velocity can be relativistic (values greather than
0.3 of the light velocity, see Fig. \ref{velocidade-trajetoria-WD}(b)) for the parameters $A=2$ and $B=1$, considered in the
presente paper. With these features, the energy density and the 
radiation force cannot be investigated via approximate methods, which require oscillatory motions with small amplitudes. To
visualize the behavior of the energy density and the radiation force, we will use next the exact approach
proposed by Cole-Schieve \cite{Cole-Schieve-1995,Cole-Schieve-2001}. 
\begin{figure}
\begin{center}
\epsfig{file=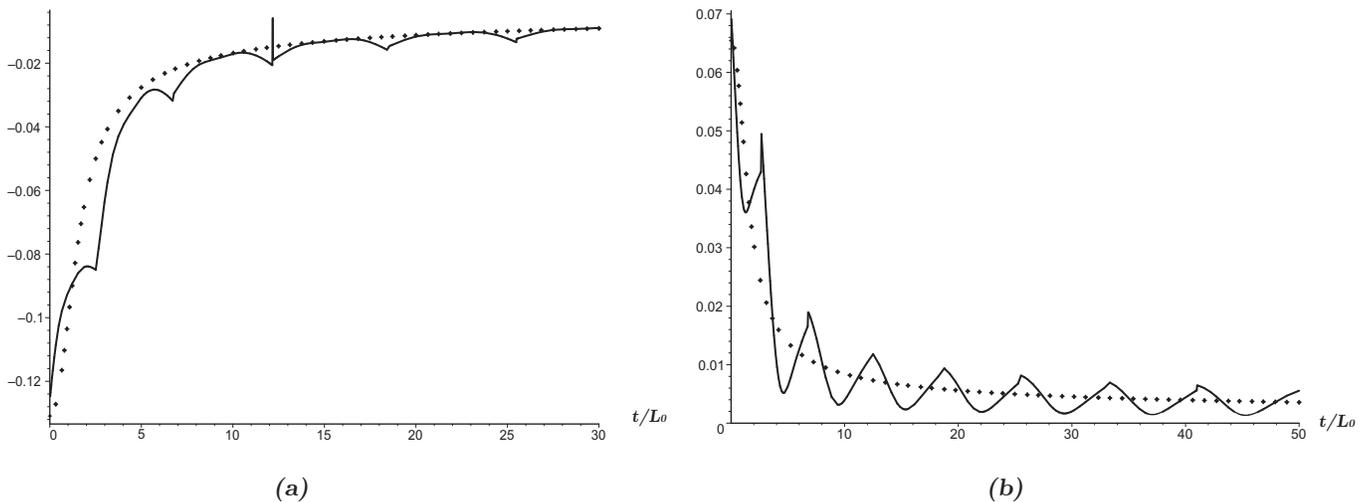,angle=-00,width=1.0 \linewidth,clip=}
\caption{\footnotesize  {\bf(a)} The quantum force acting on the moving boundary (solid line) and
the attractive static Casimir force (dotted line) for the DD or NN cases; {\bf(b)} The quantum force acting on the moving boundary
(solid line) for the DN or ND cases, and the repulsive static Casimir force (dotted line). In both cases
the vacuum was considered as the initial field state.}
\label{forca-vacuo-WD-DD}
\end{center}
\end{figure}
\begin{figure}
\begin{center}
\epsfig{file=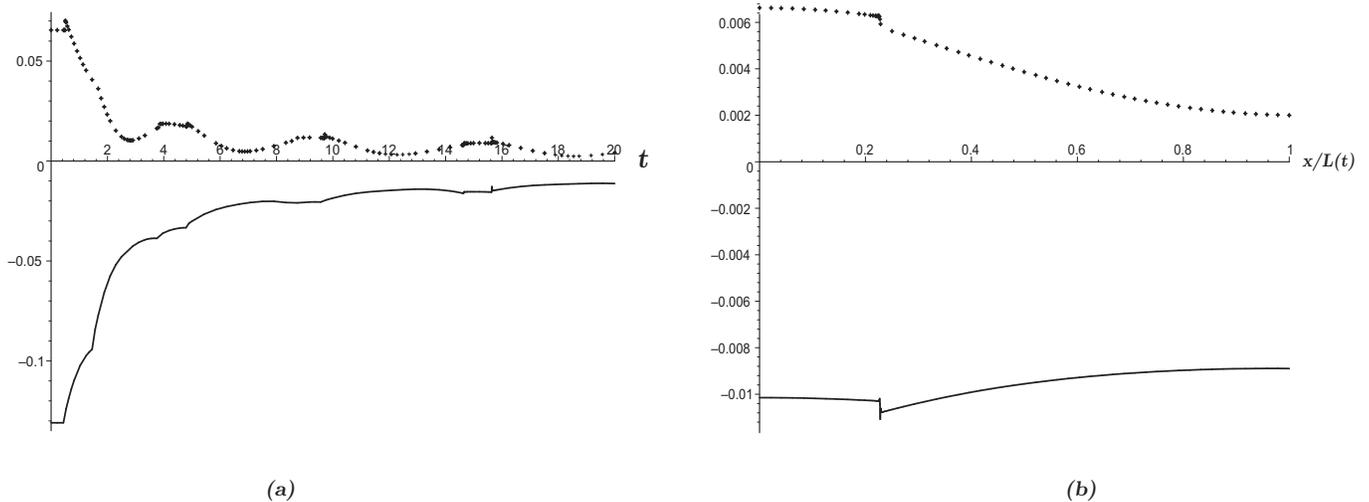,angle=-00,width=1.0 \linewidth,clip=}
\caption{\footnotesize {\bf (a)} The energy density for DD or NN cases (solid line) and also for DN or ND cases (dotted line), 
at the point $x=L_0/2$, as function of time; {\bf (b)} The energy density for for DD or NN cases (solid line) and also for DN or ND cases (dotted line),
as function of the normalized position $x/L(t)$ in the cavity, for a time $t=30L_0$. In both cases
the vacuum was considered as the initial field state.}
\label{densidade-energia}
\end{center}
\end{figure}

In Fig. \ref{forca-vacuo-WD-DD}(a) we plot, for both DD and NN cases and the vacuum as the intial
field state, the time evolution of the 
actual force acting on the moving boundary (solid line) for each position $L$, whereas the dotted
line shows the value of the static Casimir force $-\pi/[24L^2]$ which would act on the boundary if it was
static at the position $x=L$. In analogous manner, in Fig. \ref{forca-vacuo-WD-DD}(b) we plot, for the DN and ND cases, 
the time evolution of the force acting on the moving boundary (solid line), whereas the dotted
line shows the repulsive static Casimir force $\pi/[48L(t)^2]$. We see that for DD and NN, and
also for DN and ND, the radiation force acting on the moving boundaries are the same. In both
cases, as expected for this law of motion, the figures show the dynamical force approaching to the static Casimir one in the asymptotic limit $t\rightarrow \infty$. In Fig. \ref{densidade-energia} we see the energy density for several boundary conditions.
\begin{figure}
\begin{center}
\epsfig{file=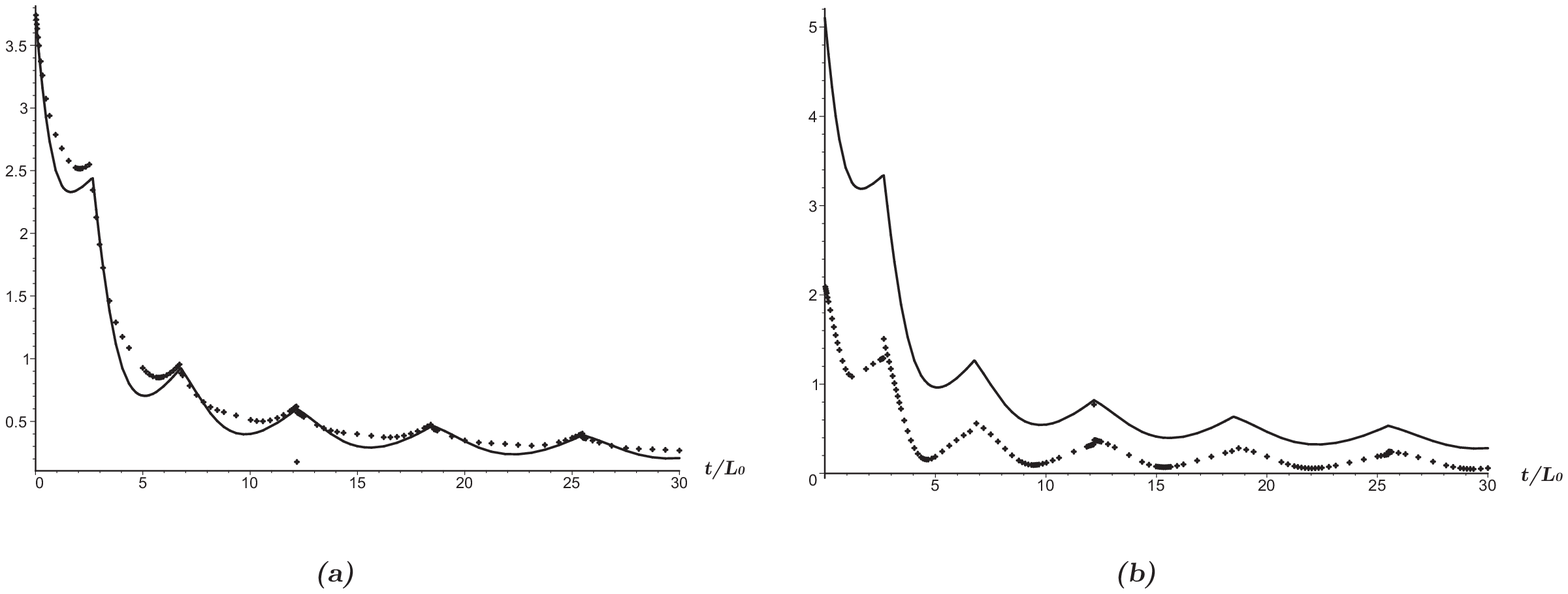,angle=-00,width=1.0 \linewidth,clip=}
\caption{\footnotesize {\bf (a)} The quantum forces ${\cal T}_{\footnotesize\;\mbox{non-vac}}(t,L(t))$
(solid line) and -30$\times$${\cal T}_{\footnotesize\;\mbox{vac}}(t,L(t))$ (dotted line) for DD or NN cases,
as function of time; {\bf (b)} The quantum forces ${\cal T}_{\footnotesize\;\mbox{non-vac}}(t,L(t))$
(solid line) and $30\times{\cal T}_{\footnotesize\;\mbox{vac}}(t,L(t))$ (dotted line) for DN or ND cases, as function of time.
In both cases the thermal bath with temperature $T=1$ was considered as the initial field state.}
\label{densidade-energia-DD-L0-sobre-2}
\end{center}
\end{figure}

Now, let us consider the thermal state as the initial field state. 
For this case we have $\langle \hat{{a}}^{\dagger}_{n}\hat{a}_{n'}\rangle=\delta_{nn'}\overline{n}(n,\beta)$ and $\langle \hat{a}_{n}\hat{a}_{n^\prime}\rangle=\langle \hat{{a}}^{\dagger}_{n}\hat{a}^{\dagger}_{n'}\rangle=0$, where $\overline{n}\left(n,\beta\right)=\left[exp(\kappa (n+\beta))-1\right]^{-1}$ and $\kappa=1/T$. In Fig.
\ref{densidade-energia-DD-L0-sobre-2} we compare the behavior of ${\cal T}_{\footnotesize\;\mbox{non-vac}}$ 
for $T=1$ and ${\cal T}_{\footnotesize\;\mbox{vac}}$. We also see that
${\cal T}_{\footnotesize\;\mbox{non-vac}}^{\footnotesize\mbox{DN}}$=${\cal T}_{\footnotesize\;\mbox{non-vac}}^{\footnotesize\mbox{ND}}$,
${\cal T}_{\footnotesize\;\mbox{non-vac}}^{\footnotesize\mbox{DD}}$=${\cal T}_{\footnotesize\;\mbox{non-vac}}^{\footnotesize\mbox{NN}}$,
and that the difference between ${\cal T}_{\footnotesize\;\mbox{non-vac}}^{\footnotesize\mbox{DD}}$ and
${\cal T}_{\footnotesize\;\mbox{non-vac}}^{\footnotesize\mbox{DN}}$ is a scale factor.

\section{Non-diagonal states: coherent and Schrödinger cat states}
\begin{figure}
\hspace{1.5cm}
\begin{center}
\epsfig{file=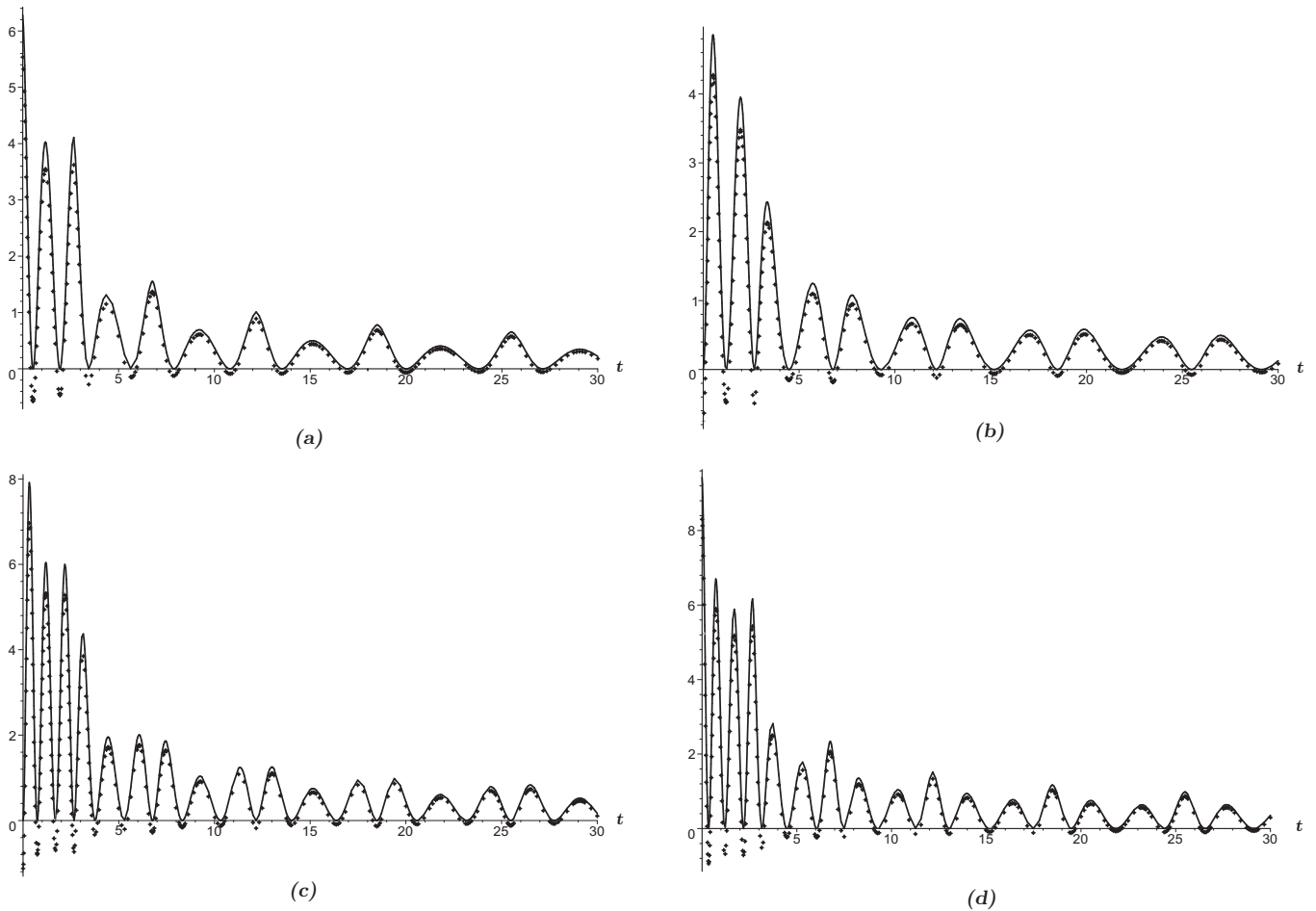,angle=00,width=1 \linewidth,clip=}
\caption{\footnotesize The non-vacuum part of the force (${\cal T}_{\footnotesize\;\mbox{non-vac}}(t,L(t))$) acting on the moving boundary
(vertical axis), as function of time.
The solid line is the coherent case, whereas the dotted line is the Schrödinger cat case. The law of motion
considered is given by Eq. (\ref{trajetoria_walker}). The parameters considered are: $|\alpha|=1$, $\theta=\phi=0$.
{\bf (a)} The DD case; {\bf (b)} The NN case; {\bf (c)} The DN case; {\bf (d)} The ND case.}
\label{forca-coerente-gato}
\end{center}
\end{figure}

%
The coherent state can be defined as an eigenstate of the annihilation operator: $\hat{a}_{n}\left|\alpha\right\rangle=\alpha\delta_{nn_{0}}\left|\alpha\right\rangle$, where 
$\alpha={\left|\alpha\right|}e^{i\theta}$ and
$n_{0}$ is related to the frequency of the excited mode \cite{Glauber}. 
The Schrödinger cat state \cite{Gato,Jeong-2002}, is defined 
as a superposition of coherent states $\left|\Psi\right\rangle=N\left(\left|\alpha\right\rangle+e^{i\phi}\left|-\alpha\right\rangle\right)$, 
where $\left|-\alpha\right\rangle$ has the same amplitude than $\left|\alpha\right\rangle$ but with
a phase shift of $\pi$, and the normalization $N$ is given by $N=(2+2e^{-2{\left|\alpha\right|}^{2}}\cos{\phi})^{-{1}/{2}}$.
When $|\alpha|$ is as small as 2, $|\langle\alpha|-\alpha\rangle|^2\approx 0$ \cite{Jeong-2005}. 
Then we expect that the cat state behaves like two coherent states.
In Fig. \ref{forca-coerente-gato} we obtain the behavior of the force acting on the moving boundary.
For $|\alpha|=1$, we can observe differences between the non-vacuum forces 
for coherent and Schrödinger cat states.
\section{Final comments}
\hspace{0.5 cm}
In the present paper we investigated the problem of the energy density inside
a non-static cavity. Since the law of motion considered here presents relativistic
velocities and also large displacement, we could not have used the usual approximate
approaches found in the literature. However, the problem can be studied
via an exact numerical approach, taking as basis the one purposed by Cole and Schieve.
In the context of this method of calculation, we 
observed that the energy density and the force on the moving mirror are not
affected if we change NN by DD boundary conditions, for vacuum or thermal initial
field states. Similar behavior is observed by changing DN by ND conditions in the context
of these diagonal states. On the other hand, the same invariance is not observed
for the non-diagonal states investigated, 
for instance, the coherent and Schrödinger cat states. For these states
each one of the cases (DD, NN, ND and DN) has a particular behavior inside the non-
static cavity.
{This work was supported by CNPq and CAPES-Brazil.}
\newpage
%

\end{document}